\documentclass[
	highlight, 
	british,
	fleqn,orivec,envcountsect,runningheads, %
	a4paper
]{llncs}
\pdfoutput=1

\makeatletter
\newcommand*{\installoption}[2][article]{
	\expandafter\newif\csname if#2\endcsname
	\@ifclasswith{#1}{#2}{
		\csname#2true\endcsname
	}{}
}
\newcommand*{\obeyoption}[2]{
	\csname if#2\endcsname\else\csname #1false\endcsname\fi
}
\newcommand*{\dominateoption}[2]{
	\csname if#2\endcsname\csname #1false\endcsname\fi
}
\makeatother

\installoption{draft}
\installoption{highlight}
\installoption{showframe}
\obeyoption{showframefalse}{draft}

\usepackage{etoolbox}
\usepackage{xspace}
\usepackage{ragged2e}

\usepackage{subfiles}
\usepackage{currfile}

\ifdefined\docroot\else
	\let\docroot\currfilepath
\fi

\newcommand\IfDocRootTF[2]{%
	\ifcurrfilepath{\docroot}{#1}{#2}%
}
\newcommand\IfDocRootT[1]{\IfDocRootTF{#1}{}}
\newcommand\IfDocRootF[1]{\IfDocRootTF{}{#1}}

\usepackage[T1]{fontenc}
\usepackage[utf8]{inputenc}
\usepackage[final]{microtype}

\ifshowframe
\AtEndPreamble{%
	\usepackage{showframe}%

}
\fi

\usepackage[usenames,dvipsnames]{xcolor}

\colorlet{type}{blue!60!black}
\colorlet{process}{red!60!black}

\colorlet{textcolor}{.}
\newcommand\usecolor[2]{\begingroup%
	\colorlet{uc@saved}{.}
	\color{#1}#2\color{uc@saved}
\endgroup}
\newcommand\cboxed[2]{\begingroup%
  \colorlet{cb@saved}{.}%
  \color{#1}%
  \boxed{\color{cb@saved}#2}%
\endgroup}

\usepackage[british]{babel}
\usepackage[all]{foreign}
\defnotforeign[wrt]{{\xperiodafter{w.r.t}}}
\defnotforeign[wl]{{\xperiodafter{w.l.o.g}}}

\usepackage{amsmath}
\usepackage{amsfonts}
\usepackage{amssymb}
\usepackage{stmaryrd}
\SetSymbolFont{stmry}{bold}{U}{stmry}{m}{n}
\usepackage{mathtools}
\usepackage{esvect}

\usepackage{proof-dashed}

\allowdisplaybreaks

\usepackage{array}
\usepackage{xtab}
\usepackage{booktabs}
\usepackage{dashrule}

\usepackage[section]{placeins}

\usepackage[inline]{enumitem}

\usepackage{hyperref}
\usepackage{nameref}

\usepackage[capitalise,nameinlink,noabbrev]{cleveref}
\makeatletter
\newcommand{\customlabel}[4][0]{%
	\protected@write\@auxout{}{\string\newlabel{#3}{{#4}{\thepage}{#4}{#3}{}}}%
	\protected@write\@auxout{}{\string\newlabel{#3@cref}{{[#2][#1][#1]#4}{\thepage}}}%
}
\makeatother

\hypersetup{
	hidelinks,
	final,
}

\spnewtheorem*{convention}{Convention}{\bfseries}{\normalfont}
\spnewtheorem*{notation}{Notation}{\bfseries}{\normalfont}

\usepackage[
	sectionbib,
	square,
	numbers,
	sort&compress
	]{natbib}
\usepackage[
	obeyDraft,   
	textwidth=3cm,
	textsize=small,
]{todonotes}

\let\phl\relax 
\let\thl\relax 
\let\nohl\relax 

\ifhighlight
\newenvironment{highlight}
	{\def\phl##1{\usecolor{process}{##1}}%
	 \def\thl##1{\usecolor{type}{##1}}%
	 \def\nohl##1{\usecolor{textcolor}{##1}}}
	{\let\phl\relax%
	 \let\thl\relax%
	 \let\nohl\relax\ignorespacesafterend}
\else
\newenvironment{highlight}{}{}
\fi

\newcommand{\m}[1]{\mathsf{#1}}

\DeclareMathOperator{\fn}{\m{fn}}
\DeclareMathOperator{\bn}{\m{bn}}

\newcommand*{\til}[1]{\tilde{#1}}

\newcommand*{\bang}{\mathord{!}}

\newcommand*{\query}{\mathord{?}}

\newcommand*{\minl}{\m{inl}}
\newcommand*{\minr}{\m{inr}}
\newcommand*{\mdisp}{\m{disp}}
\newcommand*{\muse}[1]{\m{use}\,#1}
\newcommand*{\mspawn}[1]{\m{spawn}\,#1}
\newcommand*{\sends}[2]{{#1[#2]}}
\newcommand*{\recvs}[2]{{#1(#2)}}
\newcommand*{\closes}[1]{\sends{#1}{}}
\newcommand*{\waits}[1]{\recvs{#1}{}}
\newcommand*{\inls}[1]{\sends{#1}{\minl}}
\newcommand*{\inrs}[1]{\sends{#1}{\minr}}
\newcommand*{\coinls}[1]{\recvs{#1}{\minl}}
\newcommand*{\coinrs}[1]{\recvs{#1}{\minr}}
\newcommand*{\sendtypes}[2]{\sends{#1}{#2}}
\newcommand*{\recvtypes}[2]{\recvs{#1}{#2}}
\newcommand*{\clientuses}[2]  {\sends{#1}{\muse{#2}}}
\newcommand*{\clientdisps}[1] {\sends{#1}{\mdisp}}
\newcommand*{\clientspawns}[2]{\sends{#1}{\mspawn{#2}}}
\newcommand*{\serveruses}[2]  {\recvs{#1}{\muse{#2}}}
\newcommand*{\serverdisps}[1] {\recvs{#1}{\mdisp}}
\newcommand*{\serverspawns}[2]{\recvs{#1}{\mspawn{#2}}}
\newcommand*{\forwards}[2]{{#1 \rightarrow #2}}
\newcommand{\nil}{\boldsymbol{0}}
\let\forward\forwards
\newcommand*{\pp}{\mathbin{\boldsymbol{|}}}

\newcommand*{\res}[3]{\mathop{(\boldsymbol{\nu}{#1#2})}#3}
\newcommand*{\prefixed}[2]{{#1.#2}}
\newcommand*{\send}[3]{\prefixed{\sends{#1}{#2}}{#3}}
\newcommand*{\recv}[3]{\prefixed{\recvs{#1}{#2}}{#3}}
\let\close\closes
\newcommand*{\wait}[2]{\prefixed{\waits{#1}}{#2}}
\newcommand*{\inl}[2]{\prefixed{\inls{#1}}{#2}}
\newcommand*{\inr}[2]{\prefixed{\inrs{#1}}{#2}}
\newcommand*{\Case}[3]{{#1}.\m{case}(#2,#3)}
\newcommand*{\sendtype}[3]{\prefixed{\sendtypes{#1}{#2}}{#3}}
\newcommand*{\recvtype}[3]{\prefixed{\recvtypes{#1}{#2}}{#3}}
\newcommand*{\clientdisp}[2]{\prefixed{\clientdisps{#1}}{#2}}
\newcommand*{\clientuse}[3]{\prefixed{\clientuses{#1}{#2}}{#3}}  
\newcommand*{\clientspawn}[3]{\prefixed{\clientspawns{#1}{#2}}{#3}}
\newcommand*{\server}[3]{\prefixed{\bang{#1}}{(#2)#3}}

\newcommand*{\CP}{{\upshape CP}\xspace}
\newcommand*{\CT}{{\upshape CT}\xspace}
\newcommand*{\CLL}{{\upshape CLL}\xspace}

\let\seq\vdash

\newcommand*{\dual}[1]{#1^{\bot}}

\newcommand*{\tensor}{\otimes}

\newcommand*{\parr}{\mathbin{\bindnasrepma}}

\newcommand*{\with}{\mathbin{\binampersand}}

\newcommand*{\one}{1}

\newcommand*{\zero}{0}

\newcommand*{\reducesto}{\longrightarrow}
\newcommand*{\lto}{\xrightarrow}
\newcommand{\rname}[1]{\ensuremath{\textsc{\MakeLowercase{#1}}}}
\newcommand{\rlabel}[2]{{%
	\customlabel{infrule}{#2}{#1}%
	\hypertarget{#2}{#1}}}
\crefname{infrule}{rule}{rules}
\Crefname{infrule}{Rule}{Rules}

\newcommand{\judge}[2]{\phl{#1} \seq \thl{#2}}
\newcommand{\cht}[2]{\phl{#1}\colon \thl{#2}}

\newcommand*{\lsync}[2]{\{#1,#2\}}
\newcommand{\ctlbl}[2]{\phl{#1}\colon\alignatthis\thl{#2}}
\newcommand*{\lsend}[4]{\ctlbl{\sends{#1}{#2}}{#3 \tensor #4}}
\newcommand*{\lrecv}[4]{\ctlbl{\recvs{#1}{#2}}{#3 \parr #4}}
\newcommand*{\lclose}[1]{\ctlbl{\closes{#1}}{\one}}
\newcommand*{\lwait}[1]{\ctlbl{\waits{#1}}{\bot}}
\newcommand*{\lsendtype}[2]{\ctlbl{\sendtypes{#1}{#2}}{\exists #2}}
\newcommand*{\lrecvtype}[2]{\ctlbl{\recvtypes{#1}{#2}}{\forall #2}}
\newcommand*{\linl}[3]{\ctlbl{\inls{#1}}{#2 \oplus_1 #3}}
\newcommand*{\lcoinl}[3]{\ctlbl{\coinls{#1}}{#2 \with_1 #3}}
\newcommand*{\linr}[3]{\ctlbl{\inrs{#1}}{#2 \oplus_2 #3}}
\newcommand*{\lcoinr}[3]{\ctlbl{\coinrs{#1}}{#2 \with_2 #3}}
\newcommand*{\lforward}[3]{\ctlbl{\forwards{#1}{#2}}{\ref*{rule:axiom}\,{#3}}}
\newcommand*{\lclientuse}[3]{\ctlbl{\clientuses{#1}{#2}}{\query #3}}
\newcommand*{\lserveruse}[3]{\ctlbl{\serveruses{#1}{#2}}{\bang #3}}
\newcommand*{\lclientspawn}[3]{\ctlbl{\clientspawns{#1}{#2}}{\ref*{rule:contract}\query #3}}
\newcommand*{\lserverspawn}[3]{\ctlbl{\serverspawns{#1}{#2}}{\ref*{rule:contract}\bang #3}}
\newcommand*{\lclientdisp}[2]{\ctlbl{\clientdisps{#1}}{\ref*{rule:weaken}\query #2}}
\newcommand*{\lserverdisp}[2]{\ctlbl{\serverdisps{#1}}{\ref*{rule:weaken}\bang #2}}
\newcommand{\pttrs}[3]{%
		\cboxed{black!20}{\array{c}#1\endarray}
		\xrightarrow{#2}
		\cboxed{black!20}{\array{c}#3\endarray}
}
\newcommand{\vpttrs}[4][]{%
		\array[#1]{l}
		\cboxed{black!20}{\array{c}#2\endarray}
		\\
		\quad
		\cboxed{white}{\scriptstyle\left\downarrow\cboxed{white}{\scriptstyle#3}\right.}\\
		\cboxed{black!20}{\array{c}#4\endarray}
		\endarray	
}

\begin{document}

\title{Classical Transitions}

\def\authorcid#1#2{\href{https://orcid.org/#2}{#1}}

\author{
	\authorcid{Fabrizio Montesi}{0000-0003-4666-901X}
	\and
	\authorcid{Marco Peressotti}{0000-0002-0243-0480}
}
\authorrunning{F.~Montesi and M.~Peressotti}

\institute{
	IMADA, 
	University of Southern Denmark,
	Odense,
	Denmark
 	\\
 	\email{\{fmontesi,peressotti\}@imada.sdu.dk}
}


\maketitle

\begin{abstract}\looseness=-1
We introduce the calculus of Classical Transitions (\CT), which extends the research line on the 
relationship between linear logic and processes to labelled transitions.
The key twist from previous work is registering parallelism in typing judgements, by generalising 
linear logic judgements from one sequents to many (hypersequents).
This allows us to bridge the gap between the structures of operators used as proof terms in 
previous work and those of the standard $\pi$-calculus (in particular parallel operator and 
restriction).
The proof theory of \CT allows for new proof transformations, which we show correspond to a
labelled transition system (LTS) for processes.
We prove that \CT enjoys subject reduction and progress.
\end{abstract}

\section{Introduction}
\label{sec:introduction}

Classical Processes (\CP) \cite{W14} is a process calculus inspired by the correspondence between 
the session-typed $\pi$-calculus and linear logic \cite{CP10}, where processes correspond to 
proofs, session types (communication protocols) to propositions, and communication to cut 
elimination.
Bridging process languages to linear logic paves the way to apply methods developed in one field 
to the other. This already worked for a few results, in both 
directions. For example, the proof theory of linear logic can be used to guarantee 
progress for processes \cite{CP10,W14}, and multiparty session types, originally developed for 
processes \cite{HYC16}, inspired a generalisation of the standard cut rule to the composition of
an arbitrary number of proofs, allowing for safe circular dependencies among proofs \cite{CMSY17}.

The hallmark of \CP is that the semantics of processes is given by sound proof transformations in 
Classical Linear Logic (\CLL). While this permits reusing the metatheory of linear logic ``as is'' 
to reason about process behaviour (\eg, cut elimination yields communication progress), it also 
exhibits some fundamental discrepancies with the key operators of the $\pi$-calculus 
\cite{MPW92}.

Some discrepancies are syntactic.
For example, the term for output of a linear name is $x[y].(P\pp Q)$, read ``send $y$ over $x$ and 
proceed as $P$ in parallel to $Q$''. Notice that the term constructor for output here actually 
takes $x$, $y$, $P$ and $Q$ as parameters at the same time. This discrepancy is caused by 
adopting processes as proof terms for \CLL: the typing rule for 
output (\ie, the $\tensor$ rule of \CLL) checks that the processes 
respectively implementing the behaviours of $y$ ($P$) and of $x$ ($Q$) share no resources, by 
taking two premises ($P$ and $Q$).
In general, there is no independent parallel term $P\pp Q$ in the grammar of \CP, and even if we 
added it as the $\rname{Mix}$ rule suggested in the original presentation of \CP \cite{W14}, it 
would not allow $P$ and $Q$ to communicate as in standard $\pi$-calculus. Synchronisation is 
governed instead by the restriction operator $\res xy{(P\pp Q)}$ (we use the latest syntax for \CP, 
from \cite{CLMSW16}), which links $x$ at $P$ with $y$ at $Q$ to enable communication.
Again, parallel is mixed with another operator (restriction here), but in this case it means that 
$P$ and $Q$ will communicate.

The discrepancies carry over from syntax (and typing) to semantics.
The rule for reducing an output with an input in \CP is the following.
\[
\res xy{\left(\send {x}{x'}{(P\pp Q)} \pp \recv{y}{y'}{R}\right)}
\reducesto
\res{x'}{y'}{
	\left(P  \pp
	\res{x}{y}{(Q\pp R)}
	\right)
}
\]
Notice how the rule needs to inspect the structure of the continuation of the output term ($P \pp 
Q$) to produce a typable structure for the resulting network, by nesting restrictions appropriately.

A consequence of the discrepancies is that \CP still misses a labelled transition system 
(LTS) semantics.
Keeping with our example, it is difficult to define a transition axiom for output, as in $x[y].(P 
\pp Q) \lto{x[y]} P\pp Q$, because it is not possible to type $P \pp Q$. Even if it were,
we hit another problem when attempting to recreate the reduction above using transitions. Ideally, 
we should be able to define a rule that does not inspect the structure of processes, but only their 
observables, as follows.
\[
\infer{
	\res xy{(P \pp Q)}
	\lto\tau
	\res {x'}{y'}
	{
		\res xy{(P' \pp Q')}
	}
}{
	P \lto{\sends{x}{x'}} P'
	&
	Q \lto{\recvs{y}{y'}} Q'
}
\]
However, this is not possible because the restriction term in the result is not typable in (nor is 
in the syntax of) \CP.

In this paper, we present the calculus of Classical Transitions (\CT), an attempt at mending the 
discrepancies that we discussed.
The key twist from \CP to \CT is to generalise the judgement form of \CLL from one sequent 
to collections of sequents, called hypersequents \cite{A91}.
Crucially, we use the separation of hypersequents to register the ``parallelism'' of propositions 
(as in, manipulated by separate proofs);
in particular, we interpret the composition of hypersequents as parallel composition of processes.
This allows us to redefine the typing rules of \CP such that, whenever parallelism is required, we 
can guarantee it by looking at the structure of \emph{types} (hypersequents) instead of the 
\emph{syntax} of terms. Following this principle, the adaptation of all rules in \CLL is 
straightforward.
In \CT, referring to our previous examples, the syntax for output is $x[y].P$, that for restriction 
is $\res xyP$, and that for parallel composition is $P\pp Q$, and our typing rules follow this 
structure in the expected way.

The proof theory of \CT allows for new sound proof transformations \wrt~\CP, which we show 
correspond to labelled transitions for processes, yielding an LTS semantics. We show 
that \CT enjoys subject reduction and progress (terms never get stuck, implying lack of deadlocks).
Differently from \CP, our progress result does not require any commuting conversions: actions are 
executed in place (just as in the LTS for the $\pi$-calculus), instead of being permuted inside or 
outside of parallel compositions as in \cite{W14}. Our semantics also evidences syntactically the 
explicit resource management that \CLL performs whenever server processes are 
replicated, which is hidden by ``communicating'' name substitutions in \CP.

We envision that bridging the gap that we discussed and giving an LTS semantics to \CLL proofs 
(adapted to hypersequents) will push even further the successful research line that investigates 
the relationship between linear logic and processes. 

\section{Classical Transitions}
We present Classical Transitions (\CT), a strict generalisation of the latest version of the calculus of Classical Processes \cite{CLMSW16}.

\paragraph{Processes}
In \CT, programs are processes ($P$,$Q$,$R$,\dots) that communicate using channels names ($x$,$y$,$z$,\dots). Channels represent endpoints of sessions, as in \cite{V12,CLMSW16}. Processes are given by the grammar below; some terms include types ($A$,$B$,$C$,\dots) which will be discussed afterwards.

\begin{highlight}
\medskip\par\indent
\sbox0{$\phl P, \phl Q \Coloneqq {} $} 
\sbox1{$\clientspawn{x}{x'}{P}$}
\xentrystretch{-0.1}
\begin{xtabular}{%
	>{\raggedleft$}m{\wd0}<{$}%
	>{\raggedright$\phl\begingroup}m{\wd1}<{\endgroup$}%
	>{\quad\it}l<{}}
\phl P, \phl Q \Coloneqq {} & \send xyP & output endpoint $y$ on $x$ and continue as $P$\\
\mid {} & \recv xyP & input endpoint as $y$ from $x$ and continue as $P$ \\
\mid {} & \inl xP & select left on $x$ and continue as $P$ \\
\mid {} & \inr xP & select right on $x$ and continue as $P$ \\
\mid {} & \Case{x}{P}{Q} & offer on $x$ to continue as $P$ (left) or $Q$ (right)\\
\mid {} & \sendtype{x}{A}{P} & output type $A$ on $x$ and continue as $P$\\
\mid {} & \recvtype{x}{X}{P} & input a type as $X$ in $P$ \\
\mid {} & \close x & close endpoint $x$ and terminate\\
\mid {} & \wait xP & wait for $x$ to be closed and continue as $P$\\
\mid {} & \forward{x}{y} & forward endpoint $x$ to $y$\\
\mid {} & \server{x}{y}{P} & server offering service $(y)P$ on $x$\\
\mid {} & \clientuse{x}{y}{P} & client service request \\
\mid {} & \clientspawn{x}{x'}{P} & client request spawn\\
\mid {} & \clientdisp{x}{P} & client dispose service \\
\mid {} & \res{x}{y}{P} & link endpoints $x$ and $y$ in $P$\\
\mid {} & P \pp Q & parallel composition of processes $P$ and $Q$\\
\mid {} & \nil & terminated process
\end{xtabular}
\end{highlight}
\par\medskip\noindent
We first discuss terms that are unchanged wrt \CP.
We use Wadler's convention of denoting outputs with square brackets and inputs with round 
parentheses \cite{W14}.
Term $\send{x}{y}{P}$ denotes a process that sends a fresh name $y$ over $x$ and then proceeds as $P$.
Dually, term $\recv{x}{y}{P}$ receives a name $y$ over $x$ and then proceeds as $P$.
Thus both input and output actions bind their object in continuations, as in the internal $\pi$-calculus \cite{S96}---thanks to links, it is easy to recover free output as syntactic sugar, see \cite{LM15}.
Term $\close x$ closes channel $x$, and term $\wait xP$ waits for $x$ to be closed before continuing 
as $P$.
Terms $\inl{x}{P}$ and $\inr{x}{P}$ respectively select the left and right branch of a (binary) offer available over $x$ before proceeding as $P$.
Dually, term $\Case{x}{P}{Q}$ offers over $x$ a choice between proceeding as $P$ (left branch) or 
$Q$ (right branch).
Term $\sendtype{x}{A}{P}$ sends type $A$ over $x$ and term $\recvtype{x}{X}{P}$ receives a type to 
replace $X$ with in the continuation $P$ (binding $X$ in $P$).
Term $\forward{x}{y}$ is a forwarding proxy: inputs on $x$ are forwarded as outputs on $y$ and 
\viceversa.

We now move to terms that are new or changed wrt \CP.

Term $\server{x}{y}{P}$ is a server that offers on $x$ a replicable process $P$, where $y$ is bound 
in $P$. Server channels ($x$ in the server term) are typed with the exponential connective $\bang$ 
of \CLL, which guarantees that channel $x$ can be used at will (zero, one, or many times). The 
number of times that a server channel is used is determined by the process connected to the server 
(the client).
We thus interpret the server channel as offering three possibilities, which can be selected 
from by clients (if you like, this can be seen as a variant or tagged union type).
Specifically, term $\clientuse{x}{y}{P}$ requests the server connected to $x$ to use its 
replicable process exactly once, and to continue communicating with the latter on $y$.
Term $\clientspawn{x}{x'}{P}$ requests the server on $x$ to duplicate itself and to make the new 
server accessible over $x'$.
Finally, term $\clientdisp{x}{P}$ (for ``dispose'') terminates the server---\ie, it informs it that 
it will be used zero times.

In \CP, only the action for using a server once has an explicit term (our $\clientuse{x}{y}{P}$). 
Duplication and disposal are visible only from the proof used to type a process. This yields a 
slightly unexpected reduction semantics, where a process may communicate with a server (\eg, for 
its disposal) without consuming any syntactic term (disposal is an explicit communication in \CP, 
between the proof of the client and that of the server).
We chose to make all client terms explicit in \CT, for two reasons. First, we will see that this 
allows processes to represent faithfully the structure of the proof with which they are typed, 
since now all rule applications have a corresponding term constructor. Second, when we will 
formulate the LTS semantics of \CT in the next section, we shall see that all three client 
invocations (usage, duplication, and disposal) correspond to transitions with observable actions. 
Having explicit client terms thus allows us to give transition rules in an SOS style: client 
actions will arise from syntactically corresponding terms, as usual (which would not be possible 
with the ``silent'' treatment of duplication and disposal in \CP).

A restriction term $\res{x}{y}{P}$ connects endpoints $x$ and $y$ to form a session, allowing the 
two endpoints to communicate---and binding the names $x$ and $y$ to $P$.
This term was originally introduced in \cite{V12} for the session-typed $\pi$-calculus. Later, it 
was adopted in \CP \cite{CLMSW16}, but with the arity problem discussed in the Introduction. Our 
term, instead, is exactly the same as that in \cite{V12}, which is logically reconstructed in a 
precise way for the first time here.
\CT also has the standard parallel composition term $P \pp Q$, and the terminated process term 
$\nil$.
We extend the terminology to terms that are parallel compositions of $\nil$, \ie, we say that a 
process is terminated if it is a parallel composition of $\nil$ terms.

\paragraph{Types}
There are two kinds of types in \CT: channel types (also called session types) and process types.

Channel types ($A$, $B$, $C$, \dots) are standard \CLL propositions.
They are defined by the following grammar, where $X$ ranges over atomic propositions.

\medskip 
\begin{highlight}
\sbox0{$\thl{A},\thl{B}  \Coloneqq {} $}
\sbox1{$A \tensor B$}
\begin{xtabular}{%
	>{\raggedleft$}m{\wd0}<{$}%
	>{\raggedright$\thl\begingroup}m{\wd1}<{\endgroup$}%
	>{\hspace{.9ex}\it}l<{}%
	>{\hspace{1.2ex}$}r<{$}%
	>{\raggedright$\thl\begingroup}m{\wd1}<{\endgroup$}%
	>{\hspace{.9ex}\it}l<{}}
\thl{A},\thl{B} \Coloneqq {} 
        & \thl{A \tensor B} & send $A$, proceed as $B$&
\mid {} & {A \parr B} & receive $A$, proceed as $B$\\
\mid {} & {A \oplus B} & select $A$ or $B$&
\mid {} & {A \with B} & offer $A$ or $B$\\
\mid {} & {\zero} & unit for $\oplus$&
\mid {} & {\top} & unit for $\with$\\
\mid {} & {\one} & unit for $\tensor$&
\mid {} & {\bot} & unit for $\parr$\\
\mid {} & {\query A} & client request&
\mid {} & {\bang A} & server accept\\
\mid {} & {\exists X.A} & existential&
\mid {} & {\forall X.A} & universal\\
\mid {} & {X} & atomic proposition&
\mid {} & {\dual X} & dual of atomic prop.
\end{xtabular}
\end{highlight}
\medskip\par\noindent
Types on the left-hand column are for outputs and types in the right-hand column for inputs.
Connectives on the same row are respective duals, \eg, $\tensor$ and $\parr$ are dual of each other.
We assume the standard notion of duality of \CLL, writing $\dual A$ for the dual of $A$. Duality 
proceeds homomorphically and replaced connectives with their duals, for example $\dual{(A \tensor 
B)} = \dual A \parr \dual B$.
In $\exists X.A$ and $\forall X.A$, the type variable $X$ is bound in $A$. We write $\m{ftv}(A)$ 
for the set of free type variables in $A$, and $B\{A/X\}$ to denote substitution of $A$ 
for $X$ in $B$.

Process types are \CLL hypersequents ($\Gamma$, $\Delta$, \dots), \ie, collections (multisets) of 
\CLL sequents ($\gamma$, $\delta$, \dots). Their grammar is given in the following.
\par\medskip
\begin{highlight}
\begin{tabular}{>{$}r<{$}>{$}l<{$}>{\qquad}l<{}}
\gamma,\delta \Coloneqq {} & \thl{\cht{x_1}{A_1},\dots,\cht{x_n}{A_n}} & $x_i \neq x_j$ for $i \neq j$
\\
\Gamma,\Delta \Coloneqq {} & \thl{\gamma_1 \pp \dots \pp \gamma_n} & 
$\m{n}(\gamma_i) \cap \m{n}(\gamma_j) = \emptyset$ for $i \neq j$
\end{tabular}
\end{highlight}
\medskip\par\noindent
The separator $\pp$ used in hypersequents indicates that the sequents in it are independent.
The side-conditions on the right are standard: we require channel names to be disjoint in both 
a single sequent and among all sequents in the same hypersequents. We write $\m{n}(\gamma)$ for the 
channel names in $\gamma$.
For convenience of exposition, we assume that free type variables are never shadowed by bound ones,
\eg, we assume that $X \notin \m{ftv}(\Gamma\pp\gamma)$ whenever we write $\Gamma\pp\gamma, \forall 
X.B$.
Both sequents and hypersequents allow for exchange, which we apply silently in the remainder. Likewise, we assume unit laws for empty sequents. As 
usual for linear logic, they do not allow for implicit weakening or contraction, which are managed explicitly by typing rules using exponentials.

\begin{figure}
\begin{highlight}
\begin{gather*}
  \infer[\rlabel{\rname{Ax}}{rule:axiom}]
    {\judge{\forward{x}{y}}{\cht{x}{A^\bot}, \cht{y}{A}}}
    {}
\qquad
  \infer[\rlabel{\rname{Cut}}{rule:cut}]
    {\judge{\res{x}{y}{P}}{\Gamma \pp \gamma, \delta}}
    {\judge{P}{\Gamma\pp\gamma, \cht{x}{A} \pp \delta, \cht{y}{\dual{A}}}}
\qquad
	\infer[\rlabel{\rname{Mix$_0$}}{rule:mix_0}]
		{\judge{\nil}{\cdot}}
		{}
\\
  \infer[\rlabel{\rname{Mix}}{rule:mix}]
  	{\judge{P \pp Q}{\Gamma \pp \Delta}}
	  {\judge{P}{\Gamma}&\judge{Q}{\Delta}}
\qquad
  \infer[\rlabel{\rname{$\one$}}{rule:one}]
    {\judge{\close x}{\cht{x}{\one}}}
    {}
\qquad
  \infer[\rlabel{\rname{$\bot$}}{rule:bot}]
    {\judge{\wait xP}{\gamma, \cht{x}{\bot}}}
    {\judge{P}{\gamma}}
\\
  \infer[\rlabel{\rname{$\tensor$}}{rule:tensor}]
    {\judge{\send xyP}{\gamma,\delta,\cht{x}{A \tensor B}}}
    {\judge{P}{\gamma,\cht{y}{A} \pp \delta,\cht{x}{B}}}
\qquad
  \infer[\rlabel{\rname{$\parr$}}{rule:ppar}]
    {\judge{\recv xyP}{\gamma, \cht{x}{A \parr B}}}
    {\judge{P}{\gamma, \cht{y}{A}, \cht{x}{B}}}
\\
  \infer[\rlabel{\rname{$\oplus_1$}}{rule:oplus_1}]
    {\judge{\inl xP}{\gamma, \cht{x}{A \oplus B}}}
    {\judge{P}{\gamma, \cht{x}{A}}}
\qquad
  \infer[\rlabel{\rname{$\oplus_2$}}{rule:oplus_2}]
    {\judge{\inr xP}{\gamma, \cht{x}{A \oplus B}}}
    {\judge{P}{\gamma, \cht{x}{B}}}
\\
  \infer[\rlabel{\rname{$\with$}}{rule:with}]
    {\judge{\Case xPQ}{\gamma, \cht{x}{A \with B}}}
    {\judge{P}{\gamma, \cht{x}{A}} &
    \judge{Q}{\gamma, \cht{x}{B}}}
\quad
	\infer[\rlabel{\rname{$\exists$}}{rule:exists}]
		{\judge{\sendtype xAP}{\gamma, \cht{x}{\exists X.B}}}
		{\judge{P}{\gamma, \cht{x}{B\{A/X\}}}}
\quad
	\infer[\rlabel{\rname{$\forall$}}{rule:forall}]
		{\judge{\recvtype xXP}{\gamma, \cht{x}{\forall X.B}}}
		{\judge{P}{\gamma, \cht{x}{B}}}
\\
	\infer[\rlabel{\rname{$\bang$}}{rule:!}]
		{\judge{\server xyP}{\query\gamma, \cht{x}{\bang A}}}
		{\judge{P}{\query\gamma, \cht{y}{A}}}
\qquad
	\infer[\rlabel{\rname{$\query$}}{rule:?}]
		{\judge{\clientuse xyP}{\gamma, \cht{x}{\query A}}}
		{\judge{P}{\gamma, \cht{y}{A}}}
\\
	\infer[\rlabel{\rname{W}}{rule:weaken}]
		{\judge{\clientdisp xP}{\gamma, \cht{x}{\query A}}}
		{\judge{P}{\gamma}}
\qquad
	\infer[\rlabel{\rname{C}}{rule:contract}]
		{\judge{\clientspawn xyP}{\gamma, \cht{x}{\query A}}}
		{\judge{P}{\gamma, \cht{x}{\query A}, \cht{y}{\query A}}}
\end{gather*}
\end{highlight}
\caption{Classical transitions, typing rules.}
\label{fig:typing}
\end{figure}

\paragraph{Typing}
Typing judgements in \CT have the form $\judge{P}{\Gamma}$ and read ``process $P$ uses channels 
according to $\Gamma$''. We omit empty (hyper)sequents. We say that a process $P$ is 
\emph{well-typed} whenever $\judge{P}{\Gamma}$ for some hypersequent $\Gamma$. The rules for 
deriving typing judgements are displayed in \cref{fig:typing}.

Typing rules associate types to channels by looking at how channels are used in process terms. 
Rule selection is structural on the syntax of processes, in the sense that it 
depends only on the outermost constructor of a process term.
The typing rules of \CT are those of \CLL, adapted from sequents to hypersequents as expected 
\cite{A91}.
The key twists that we introduce are the structural \cref{rule:mix,rule:cut}.
\Cref{rule:mix} types the parallel composition of two processes, by combining their types as a
hypersequent. Previous presentations of \cref{rule:mix} (\eg, \cite{W14}) do not use 
hypersequents, thus losing the information that the resources in the two premises of the rule are 
independent.
This information is crucial to our reformulation of \cref{rule:cut}, which types a restriction 
connecting endpoints $x$ and $y$ by requiring that the types of 
$x$ and $y$ are respective duals (as usual in \CLL) and are used by parallel components of the 
process (new in \CT).
The latter condition, which we can check thanks to hypersequents, makes the rule sound 
without having to inspect the structure of the restricted process.
By comparison, the standard cut rule of linear logic requires two separate \emph{proofs} as 
premises, yielding the restriction term constructor $\res{x}{y}(P \pp Q)$ that we 
discussed in the Introduction.
Our \cref{rule:tensor} is reformulated from \CLL using the same intuition for \cref{rule:cut} 
(the original rule requires two separate proofs for $A$ and $B$ respectively). This yields a 
logical reconstruction of the expected output term from the internal $\pi$-calculus \cite{S96}.

\Cref{rule:axiom} types a forwarding proxy between endpoints $x$ and $y$ by requiring 
that the types of $x$ and $y$ are respective duals. This ensures that any message on $x$ 
can be safely forwarded to $y$, and \viceversa. All rules for typing channels enforce linear usage,
aside from client requests (typed with the exponential connective $\query$), for which contraction 
and weakening are allowed.
Contraction (\cref{rule:contract}) allows for multiple client requests for the same server channel, 
and weakening (\cref{rule:weaken}) for clients that do not use a server. Thus, \CT exposes 
syntactically that \CLL yields a calculus where servers are resources managed explicitly by 
clients.

All other rules are standard. \Cref{rule:mix_0} was introduced to \CP in \cite{A17}.

\Cref{thm:sequent} below formalises that all sequents in a provable hypersequent are 
independent, in the sense that they are independently provable.
 endpoints.
\begin{proposition}
	\label{thm:sequent}
	If $\judge{}{\nohl{\Gamma \pp \gamma}}$ then, $\judge{}{\nohl\gamma}$.
\end{proposition}
Intuitively, this confirms that the parallel composition of sequents in hypersequents denotes 
non-interference. Different sequents can indeed interact only when connected by rule \rname{Cut}, 
which then merges the interacting sequents together (since they now depend on each other).

\Cref{thm:ct-extends-cp} below states that syntax and typing of \CT 
form a strict generalisation of \CP. The proof theory itself is a strict extension of \CLL since, 
\eg, $\judge{}{\nohl{\one \tensor \bot}}$ is provable in \CT but not in \CLL.
\begin{proposition}
	\label{thm:ct-extends-cp}
	If $\judge{\nohl P}{\nohl \gamma}$ in \CP then $\judge{\nohl P}{\nohl \gamma}$ in \CT but not \viceversa.
\end{proposition}

\section{Semantics}

\begin{figure}
\begin{highlight}
\renewcommand*{\ctlbl}[2]{\phl{#1}} 
\begingroup
Labels:
\medskip\par\quad
\renewcommand{\arraystretch}{1.2}
\begin{xtabular}{
	>{$}l<{$}
	>{\hspace{.8ex}\it}l<{\hspace{1.2ex}}
	>{$}l<{$}
	>{\hspace{.8ex}\it}l<{}
}
\lclose{x} & close $x$ & 
\lwait{y} & wait for $y$ to be closed\\
\lsend{x}{x'}{A}{B} & send $x'$ on $x$ &
\lrecv{y}{y'}{\dual{A}}{\dual{B}} & receive $y'$ on $y$\\
\linl{x}{A}{B} & send  select left &
\lcoinl{y}{\dual{A}}{\dual{B}} & receive select left\\
\linr{x}{A}{B} & send  select right  &
\lcoinr{y}{\dual{A}}{\dual{B}} & receive select right\\
\lsendtype{x}{A} & send type $A$ on $x$ &
\lrecvtype{y}{A} & receive type $A$ on $y$\\
\lclientuse{x}{x'}{A} & open session on $x$ as $x'$ &
\lserveruse{y}{y'}{A} & accept session on $y$ as $y'$\\
\lclientspawn{x}{x'}{A} & request spawn as $x'$& 
\lserverspawn{y}{y'}{A} & receive spawn as $y'$\\
\lclientdisp{x}{A} & request dispose& 
\lserverdisp{y}{A} & receive dispose
\end{xtabular}
\endgroup

\par\bigskip
Transitions:
\begin{gather*}
	\phl{\close x} \xrightarrow{\lclose x} \phl{\nil}
\qquad
	\phl{\wait xP} \xrightarrow{\lwait x} \phl{P}
\qquad
	\phl{\send x{x'} P} \xrightarrow{\lsend x{x'}AB} \phl{P}
\qquad
	\phl{\recv x{x'} P} \xrightarrow{\lrecv x{x'}AB} \phl{P}
\\
	\phl{\forward xy} \xrightarrow{\lforward xyA} \phl{\nil}
\qquad
	\phl{\inl xP} \xrightarrow{\linl xAB} \phl{P}
\qquad
	\phl{\Case xPQ} \xrightarrow{\lcoinl xAB} \phl{P}
\qquad
	\phl{\inr xP} \xrightarrow{\linr xAB} \phl{P}
\\
	\phl{\Case xPQ} \xrightarrow{\lcoinr xAB} \phl{Q}
\qquad
	\phl{\sendtype xAP} \xrightarrow{\lsendtype xA} \phl{P}
\qquad
	\phl{\recvtype xXP} \xrightarrow{\lrecvtype xA} \phl{P\{A/X\}}
\\
	\phl{\clientuse x{x'}P} \xrightarrow{\lclientuse x{x'}A} \phl{P}
\quad
	\phl{\server x{y}P} \xrightarrow{\lserveruse x{y}A} \phl{P}
\quad
	\phl{\clientdisp xP} \xrightarrow{\lclientdisp xA} \phl{P}
\quad
	\phl{\server x{y}P} \xrightarrow{\lserverdisp xA} \phl{\nil}
\\
	\phl{\clientspawn x{x'}P} \xrightarrow{\lclientspawn x{x'}A} \phl{P}
\qquad
	\phl{\server x{y}P} \xrightarrow{\lserverspawn x{x'}A} \phl{\server{x}{y}{P} \pp \server{x'}{y}{P'}}
\\
	\infer[\rlabel{\ref{rule:mix-l}}{rule:p-mix-l}]
		{\phl{P \pp Q} \xrightarrow{\phl{\alpha}} \phl{R \pp Q}}
		{\phl{P} \xrightarrow{\phl{\alpha}} \phl{R}
		&\scriptstyle\phl{\bn(\alpha) \cap \fn(Q) = \emptyset}}
\qquad
	\infer[\rlabel{\ref{rule:mix-r}}{rule:p-mix-r}]
		{\phl{P \pp Q} \xrightarrow{\phl{\alpha}} \phl{P \pp Q'}}
		{\phl{Q} \xrightarrow{\phl{\alpha}} \phl{Q'}
		& \scriptstyle\phl{\bn(\alpha) \cap \fn(P) = \emptyset}}
\\
	\infer[\rlabel{\ref{rule:mix-sync}}{rule:p-mix-sync}]
		{\phl{P \pp Q} \xrightarrow{\lsync{\phl{\alpha}}{\phl{\dual{\alpha}}}} \phl{R \pp S}}
		{\phl{P} \xrightarrow{\alpha} \phl{R}
		&\phl{Q} \xrightarrow{\dual{\alpha}} \phl{S}}
\qquad
	\infer[\rlabel{\ref{rule:cut-res}}{rule:p-cut-res}]
		{\phl{\res{x}{y}{P}} \xrightarrow{\phl{\alpha}} \phl{\res{x}{y}{Q}}}
		{\phl{P} \xrightarrow{\alpha} \phl{Q}&\phl{x,y \notin \m{n}(\alpha)}}
\\
	\infer[\rlabel{\ref{rule:cut-one}}{rule:p-cut-one}]
		{\phl{\res{x}{y}{P}} \xrightarrow{\phl{\tau}} \phl{Q}}
		{\phl{P} \xrightarrow{\lsync{\lclose x}{\lwait y}} \phl{Q}}
\qquad
	\infer[\rlabel{\ref{rule:cut-tensor}}{rule:p-cut-tensor}]
		{\phl{\res{x}{y}{P}} \xrightarrow{\phl{\tau}} \phl{\res{x}{y}{\res{x'}{y'}{Q}}}}
		{\phl{P} \xrightarrow{\lsync{\lsend x{x'}AB}{\lrecv y{y'}{\dual A}{\dual B}}} \phl{Q}}
\\
	\infer[\rlabel{\ref{rule:cut-oplus_1}}{rule:p-cut-oplus_1}]
		{\phl{\res{x}{y}{P}} \xrightarrow{\phl{\tau}} \phl{\res{x}{y}{Q}}}
		{\phl{P} \xrightarrow{\lsync{\linl xAB}{\lcoinl y{\dual A}{\dual B}}} \phl{Q}}
\qquad
	\infer[\rlabel{\ref{rule:cut-oplus_2}}{rule:p-cut-oplus_2}]
		{\phl{\res{x}{y}{P}} \xrightarrow{\phl{\tau}} \phl{\res{x}{y}{Q}}}
		{\phl{P} \xrightarrow{\lsync{\linr xAB}{\lcoinr y{\dual A}{\dual B}}} \phl{Q}}
\\
	\infer[\rlabel{\ref{rule:cut-?}}{rule:p-cut-?}]
		{\phl{\res{x}{y}{P}} \xrightarrow{\phl{\tau}} \phl{\res{x'}{y'}{Q}}}
		{\phl{P} \xrightarrow{\lsync{\lclientuse x{x'}A}{\lserveruse y{y'}{\dual A}}} \phl{Q}}
\qquad
	\infer[\rlabel{\ref{rule:cut-weaken}}{rule:p-cut-weaken}]
		{\phl{\res{x}{y}{P}} \xrightarrow{\phl{\tau}} 
		 \phl{\clientdisp{z_1}{\cdots\clientdisp{z_n}{Q}}}}
		{\phl{P} \xrightarrow{\lsync{\lclientdisp xA}{\lserverdisp y{\dual A}}} \phl{Q}
		&\phl{\scriptstyle\{z_1,\dots,z_n\} = \fn(P) \setminus \fn(Q)}}
\\
	\infer[\rlabel{\ref{rule:cut-contract}}{rule:p-cut-contract}]
		{\phl{\res{x}{y}{P}} \xrightarrow{\phl{\tau}} \phl{\clientspawn{z_1}{z_1'}{\cdots\clientspawn{z_n}{z_n'}{\res{x}{y}{\res{x'}{y'}{Q}}}}}}
		{\phl{P} \xrightarrow{\lsync{\lclientspawn x{x'}A}{\lserverspawn y{y'}{\dual A}}} \phl{Q}
		&\phl{\scriptstyle\{z_1,\dots,z_n\} = \fn(Q) \setminus \fn(P)}}
\\
	\infer[\rlabel{\ref{rule:cut-exists}}{rule:p-cut-exists}]
		{\phl{\res{x}{y}{P}} \xrightarrow{\phl{\tau}} \phl{\res{x}{y}{{Q}}}}
		{\phl{P} \xrightarrow{\lsync{\lsendtype x{A}}{\lrecv y{A}{\dual A}{\dual B}}} \phl{Q}}
\quad
	\infer[\rlabel{\ref{rule:cut-forward}}{rule:p-cut-forward}]
		{\phl{\res{y}{z}{P}} \xrightarrow{\tau} \phl{Q\{x/z\}}}
		{\phl{P} \xrightarrow{\lforward{x}{y}{A}} \phl{Q}}
\quad
	\infer[\rlabel{\ref{rule:cut-backward}}{rule:p-cut-backward}]
		{\phl{\res{w}{x}{P}} \xrightarrow{\tau} \phl{Q\{y/w\}}}
		{\phl{P} \xrightarrow{\lforward{x}{y}{A}} \phl{Q}}	
\end{gather*}
\end{highlight}
\caption{Classical Transitions, process labelled transition system.}
\label{fig:ct-sos}
\end{figure}

We now move to defining a semantics for \CT in terms of a labelled transition system (LTS).
The key novelty of our approach is viewing proofs as states of the LTS, and proof transformations 
as transitions. More specifically, we will show that the proof theory of \CT can be given a
labelled semantics in the SOS style \cite{P04}, by viewing:
\begin{itemize}
	\item inference rules as operations of a (sorted) signature;
	\item proofs as terms generated by this signature;
	\item (labelled) proof transformations as (labelled) transitions;
	\item and a specification of rules for deriving proof transformations as an SOS specification.
\end{itemize}
Then, a semantics for \CT processes in terms of an SOS specification is obtained simply by 
reading off how the SOS specification of proof transformations manipulate the processes that they 
type.

We illustrate the intuition behind the LTS for proof transformations.
Consider the proof for a judgement $\judge{\nohl{\wait xP}}{\nohl{\gamma, 
\cht{\nohl x}{\nohl\bot}}}$. By the
strict correspondence between term constructors and typing rules, the proof necessarily has the 
following shape.
\begin{highlight}
\[
\infer[\ref{rule:bot}]
	    {\judge{\wait xP}{\gamma, \cht{x}{\bot}}}
	    {\judge{P}{\gamma}}
\]
\end{highlight}
We can view \cref{rule:bot} as the outermost operation used in the proof.
Then, the proof of $\judge{\nohl P}{\nohl{\gamma}}$ is an argument of the 
operation, which is also parametric in channel $x$. This corresponds to the term 
constructor $\wait{x}{(-)}$ in the syntax of \CT processes---which in this case takes $P$ as 
parameter, \ie, the term corresponding to the proof of the premise.
Thus, this operation is the proof equivalent of the term constructor $\wait{x}{(-)}$ in the syntax 
of \CT processes, which denotes an observable action. Term constructors like this, also called 
action prefixes, are typically 
assigned a transition rule in process calculi. Therefore, this correspondence points at the 
transition axiom below (we box proofs for readability).
\begin{highlight}\[
\pttrs
	{\infer[\ref{rule:bot}]
	    {\judge{\wait xP}{\gamma, \cht{x}{\bot}}}
	    {\judge{P}{\gamma}}}
	{\lwait{x}}
	{\judge{P}{\gamma}}	
\]\end{highlight}
The label identifies the prefix constructor (\ie, rule name and parameter) and its syntax is 
inspired to common syntax for labels of action prefixes in process calculi.
By reading proof terms off the rule above we obtain the axiom below. 
\begin{highlight}\[
	\phl{\wait xP} \xrightarrow{\phl{\waits x}} \phl{P}
\]\end{highlight}
This axiom defines the semantics of the constructor $\wait{y}{(-)}$ as one would expect.

Following this methodology, we derive an LTS for proofs in \CT, and reading its process part we 
obtain the LTS of \CT processes given in \cref{fig:ct-sos}. 
We discuss the transition rules in the remainder of this section, by discussing the proof 
transformations that they originate from.

\subsection{Multiplicatives and \rname{Mix}}
The multiplicative fragment of \CT is formed by the 
\cref{rule:tensor,,rule:ppar,,rule:one,,rule:bot} together with the structural 
\cref{rule:mix,,rule:mix_0,,rule:cut}. 
Observe that rules from the first group have the ``action prefix'' form described above (we stretch 
the definition by regarding $\judge{\close x}{\cht{x}{\one}}$ as $\judge{\prefixed{\close 
x}{\nil}}{\cht{x}{\one}}$).
The corresponding axioms are given below.
\begin{highlight}
\[
\pttrs
	{\infer[\ref{rule:tensor}]
    {\judge{\send x{x'} P}{\gamma,\delta,\cht{x}{A \tensor B}}}
    {\judge{P}{\gamma,\cht{x'}{A} \pp \delta,\cht{x}{B}}}}
	{\lsend{x}{x'}{A}{B}}
	{\judge{P}{\gamma,\cht{x'}{A} \pp \delta,\cht{x}{B}}}
\]\[
\pttrs
	{\infer[\ref{rule:ppar}]
    {\judge{\recv y{y'} P}{\gamma, \cht{y}{A \parr B}}}
    {\judge{P}{\gamma, \cht{y'}{A}, \cht{y}{B}}}}
	{\lrecv{y}{y'}{A}{B}}
	{\judge{P}{\gamma, \cht{y'}{A}, \cht{y}{B}}}
\]\[
\pttrs
	{\infer[\ref{rule:one}]
		{\judge{\close x}{\cht{x}{\one}}}
	  {\phantom{\pp}}}
	{\lclose{x}}
	{\judge{\nil}{\cdot}}
\]
\end{highlight}
We extend the notion of duality from types to labels; for a label $\alpha$ we write $\dual{\alpha}$ for any label that describes an action that is dual to that of $\alpha$. Observe that label duality is not an involution but a binary relation: $\lsend{x}{x'}{A}{B}$ and $\lrecv{y}{y'}{\dual A}{\dual B}$ are termed dual regardless of the name parameters, likewise for $\lclose{x}$ and $\lwait{y}$.

The derivation
rules associated to \cref{rule:mix} are listed below. 
\begin{highlight}
\[
\def\regh{\vphantom{\Gamma'\pp}}
\infer[\rlabel{\rname{Par$_1$}}{rule:mix-l}]
	{\pttrs
	  {\infer[\ref{rule:mix}]
	  	{\judge{P \pp Q}{\Gamma \pp \Delta}\regh}
		  {\judge{P}{\Gamma}&\judge{Q}{\Delta}\regh}}
	  {\alpha}
	  {\infer[\ref{rule:mix}]
	  	{\judge{P' \pp Q}{\Gamma' \pp \Delta}\regh}
		  {\judge{P'}{\Gamma'}&\judge{Q}{\Delta}\regh}}}
	{\pttrs{\judge{P}{\Gamma}\regh}{\alpha}{\judge{P'}{\Gamma'}\regh} & 
	 \phl{\bn(\alpha) \cap \fn(Q) = \emptyset}}
\]\[
\def\regh{\vphantom{\pp\Delta'}}
\infer[\rlabel{\rname{Par$_2$}}{rule:mix-r}]
	{\pttrs
	  {\infer[\ref{rule:mix}]
	  	{\judge{P \pp Q}{\Gamma \pp \Delta}\regh}
		  {\judge{P}{\Gamma}&\judge{Q}{\Delta}\regh}}
	  {\alpha}
	  {\infer[\ref{rule:mix}]
	  	{\judge{P \pp Q'}{\Gamma \pp \Delta'}\regh}		  {\judge{P}{\Gamma}&\judge{Q'}{\Delta'}\regh}}}
	{\pttrs{\judge{Q}{\Delta}\regh}{\alpha}{\judge{Q'}{\Delta'}\regh} & 
 		 \phl{\bn(\alpha) \cap \fn(P) = \emptyset}}
\]\[
\def\regh{\vphantom{\Gamma'\pp\Delta'}}
\infer[\rlabel{\rname{Syn}}{rule:mix-sync}]
	{\pttrs
	  {\infer[\ref{rule:mix}]
	  	{\judge{P \pp Q}{\Gamma \pp \Delta}\regh}
		  {\judge{P}{\Gamma}&\judge{Q}{\Delta}\regh}}
	  {\lsync{\alpha}{\dual{\alpha}}}
	  {\infer[\ref{rule:mix}]
	  	{\judge{P' \pp Q'}{\Gamma' \pp \Delta'}\regh}
		  {\judge{P'}{\Gamma'}&\judge{Q'}{\Delta'}\regh}}}
	{\pttrs{\judge{P}{\Gamma}\regh}{\alpha}{\judge{P'}{\Gamma'}\regh}
	&
	\pttrs{\judge{Q}{\Delta}\regh}{\dual{\alpha}}{\judge{Q'}{\Delta'}\regh}
	}
\]
\end{highlight}
\Cref{rule:mix-l,rule:mix-r} transform one of the two parallel components composed by 
\cref{rule:mix} given that the transformation preserves non-interference in the result 
(disjointness of names). This condition follows from well-formedness of hypersequents $\Gamma \pp 
\Delta$, $\Gamma' \pp \Delta$, $\Gamma \pp \Delta'$ but must be explicitly listed as a premise if we 
read off only the process part of the rule, yielding exactly the process transition that one would 
expect for internal $\pi$-calculus.
\Cref{rule:mix-sync} pairs dual transformations of parallel components.

Alternatively to \cref{rule:mix-sync}, one may combine sets of all transformations instead of 
just duals, delegating pairing to transition rules for $\ref{rule:cut}$. This would yield 
a ``true concurrency'' interpretation of \CT instead of the standard semantics, 
which we leave to future work.

Communication under \ref{rule:cut} is modelled by transitions derived with the rules below, one for 
each type of dual labels.
\begin{highlight}
\[
\infer[\rlabel{\ref*{rule:one}\ref*{rule:bot}}{rule:cut-one}]
	{\pttrs
	  {\infer[\ref{rule:cut}] 
     {\judge{\res{x}{y}{P}}{\Gamma \pp \gamma}}
     {\judge{P}{\Gamma\pp \cht{x}{\one} \pp \gamma, \cht{y}{\bot}}}}
	  {\tau}
	  {\judge{P'}{\Gamma \pp \gamma}}
  }{\pttrs
		{\judge{P}{\Gamma\pp \cht{x}{\one} \pp \gamma, \cht{y}{\bot}}}
		{\lsync{\lclose x}{\lwait y}}
		{\judge{P'}{\Gamma \pp \gamma}}}
\]
\[
\infer[\rlabel{\ref*{rule:tensor}\ref*{rule:ppar}}{rule:cut-tensor}]
	{\vpttrs
		{\infer[\ref{rule:cut}] 
		{\judge{\res{x}{y}{P}}{\Gamma \pp \gamma,\delta, \varepsilon}}
		{\judge{P}{\Gamma\pp\gamma,\delta, \cht{x}{A \tensor B} \pp \varepsilon, \cht{y}{\dual{A} \parr \dual{B}}}}}
	{\tau}
	{\infer[\ref{rule:cut}] 
		{\judge{\res{x}{y}{\res{x'}{y'}{P'}}}{\Gamma \pp \gamma, \delta, \varepsilon}}
		{\infer[\ref{rule:cut}] 
			{\judge{\res{x'}{y'}{P'}}{\Gamma \pp \gamma,\cht{x}{B} \pp \delta, \varepsilon,\cht{y}{\dual{B}}}}
			{\judge{P'}{\Gamma \pp \gamma, \cht{x}{B} \pp \delta,\cht{x'}{A} \pp \varepsilon, \cht{y}{\dual{B}}, \cht{y'}{\dual{A}}}}}}
	}{\vpttrs
		{\judge{P}{\Gamma\pp\gamma,\delta, \cht{x}{A \tensor B} \pp \varepsilon, \cht{y}{\dual{A} \parr \dual{B}}}}
		{\lsync{\lsend x{x'}AB}{\lrecv y{y'}{\dual{A}}{\dual{B}}}}
		{\judge{P'}{\Gamma \pp \gamma, \cht{x}{B} \pp \delta,\cht{x'}{A} \pp \varepsilon, \cht{y}{\dual{B}}, \cht{y'}{\dual{A}}}}}
\]
\end{highlight}
These transformations do not interact with the context nor have any effect on the type and are hence labelled with $\tau$ as common for process calculi.
Moreover, they correspond to cut elimination in \CP \cite[Fig.~3]{W14}.
Finally, unrestricted actions are modelled by simply propagating transitions as formalised by the 
rule below.
\begin{highlight}
\[\infer[\rlabel{\rname{Res}}{rule:cut-res}]
{\pttrs
	{\infer[\ref{rule:cut}] 
		{\judge{\res{x}{y}{P}}{\Gamma \pp \gamma,\delta}}
		{\judge{P}{\Gamma \pp \gamma,\cht{x}{A} \pp \delta,\cht{y}{\dual{A}}}}}
	{\alpha}
	{\infer[\ref{rule:cut}] 
		{\judge{\res{x}{y}{P'}}{\Gamma' \pp \gamma',\delta'}}
		{\judge{P}{\Gamma' \pp \gamma',\cht{x}{A} \pp \delta',\cht{y}{\dual{A}}}}}}
{\pttrs
	{\judge{P}{\Gamma \pp \gamma,\cht{x}{A} \pp \delta,\cht{y}{\dual{A}}}}
	{\alpha}
	{\judge{P'}{\Gamma' \pp \gamma',\cht{x}{A} \pp \delta',\cht{y}{\dual{A}}}}
&\phl{x,y \notin \m{n}(\alpha)}}
\]
\end{highlight}

\subsection{Additives}
The transition rules modelling (left) selection and choice are given below and are obtained 
following the reasoning discussed above. 
\begin{highlight}
\[
\pttrs
	{\infer[\ref{rule:oplus_1}]
		{\judge{\inl xP}{\gamma, \cht{x}{A \oplus B}}}
		{\judge{P}{\gamma, \cht{x}{A}}}}
	{\linl{x}{A}{B}}
	{\judge{P}{\gamma, \cht{x}{A}}}
\]
\[
\pttrs
	{\infer[\ref{rule:with}]
		{\judge{\Case yPQ}{\gamma, \cht{y}{A \with B}}}
		{\judge{P}{\gamma, \cht{y}{A}} &
		\judge{Q}{\gamma, \cht{y}{B}}}}
	{\lcoinl{y}{A}{B}}
	{\judge{P}{\gamma, \cht{y}{A}}}
\]
\[
\infer[\rlabel{\ref*{rule:oplus_1}\ref*{rule:with}}{rule:cut-oplus_1}]
	{\vpttrs
	  {\infer[\ref{rule:cut}] 
     {\judge{\res{x}{y}{P}}{\Gamma \pp \gamma, \delta}}
     {\judge{P}{\Gamma\pp\gamma,\cht{x}{A \tensor B} \pp \delta, \cht{y}{\dual{A} \parr \dual{B}}}}}
	  {\tau}
	  {\infer[\ref{rule:cut}] 
     {\judge{\res{x}{y}{P}'}{\Gamma \pp \gamma, \delta}}
     {\judge{P'}{\Gamma \pp \gamma, \cht{x}{A} \pp \delta, \cht{y}{\dual{A}}}}}
  }{\vpttrs
		{\judge{P}{\Gamma\pp\gamma, \cht{x}{A \oplus B} \pp \delta, \cht{y}{\dual{A} \with \dual{B}}}}
				{\lsync{\linl xAB}{\lcoinl y{\dual A}{\dual B}}}
		{\judge{P'}{\Gamma \pp \gamma, \cht{x}{A} \pp \delta, \cht{y}{\dual{A}}}}}
\]
\end{highlight}
Rules for right selection are symmetric.
\begin{highlight}
\[
\pttrs
	{\infer[\ref{rule:oplus_2}]
		{\judge{\inr xP}{\gamma, \cht{x}{A \oplus B}}}
		{\judge{P}{\gamma, \cht{x}{B}}}}
	{\linr{x}{A}{B}}
	{\judge{P}{\gamma, \cht{x}{B}}}
\]\[
\pttrs
	{\infer[\ref{rule:with}]
		{\judge{\Case yPQ}{\gamma, \cht{y}{A \with B}}}
		{\judge{P}{\gamma, \cht{y}{A}} &
		\judge{Q}{\gamma, \cht{y}{B}}}}
	{\lcoinr{y}{A}{B}}
	{\judge{Q}{\gamma, \cht{y}{B}}}
\]\[
\infer[\rlabel{\ref*{rule:oplus_2}\ref*{rule:with}}{rule:cut-oplus_2}]
	{\vpttrs
	  {\infer[\ref{rule:cut}] 
     {\judge{\res{x}{y}{P}}{\Gamma \pp \gamma, \delta}}
     {\judge{P}{\Gamma\pp\gamma, \cht{x}{A \tensor B} \pp \delta, \cht{y}{\dual{A} \parr \dual{B}}}}}
	  {\tau}
	  {\infer[\ref{rule:cut}] 
     {\judge{\res{x}{y}{P}'}{\Gamma \pp \gamma, \delta}}
     {\judge{P'}{\Gamma \pp \gamma, \cht{x}{B} \pp \delta, \cht{y}{\dual{B}}}}}
  }{\vpttrs
		{\judge{P}{\Gamma\pp\gamma, \cht{x}{A \oplus B} \pp \delta, \cht{y}{\dual{A} \with \dual{B}}}}
		{\lsync{\linr xAB}{\lcoinr y{\dual A}{\dual B}}}
		{\judge{P'}{\Gamma \pp \gamma, \cht{x}{B} \pp \delta, \cht{y}{\dual{B}}}}}
\]
\end{highlight}
\subsection{Links}
There is one transition for \ref{rule:axiom}, defined by the axiom below (akin to the axiom 
for \ref{rule:one}).
\begin{highlight}
\[
\pttrs
	{\infer[\ref{rule:axiom}]
		{\judge{\forward{x}{y}}{\cht{x}{\dual{A}}, \cht{y}{A}}}
		{\phantom{\pp}}}
	{\lforward{x}{y}{A}}
	{\judge{\nil}{\cdot}}
\]
\end{highlight}
\Cref{rule:cut-forward,rule:cut-backward} below correspond to cuts of dual uses of \cref{rule:axiom}.
\begin{highlight}
\[
\infer[\rlabel{\rname{\ref*{rule:axiom}$_1$}}{rule:cut-forward}]
	{\pttrs
	  {\infer[\ref{rule:cut}] 
     {\judge{\res{y}{z}{P}}{\Gamma \pp \gamma, \cht{x}{\dual{A}}}}
     {\judge{P}{\Gamma\pp \cht{x}{\dual{A}},\cht{y}{A} \pp \gamma, \cht{z}{\dual{A}}}}}
	  {\tau}
	  {\judge{Q\{x/z\}}{\Gamma \pp \gamma, \cht{x}{\dual{A}}}}
  }{\pttrs
		{\judge{P}{\Gamma\pp \cht{x}{\dual{A}},\cht{y}{A} \pp \gamma, \cht{z}{\dual{A}}}}
		{\lforward{x}{y}{A}}
		{\judge{Q}{\Gamma \pp \gamma,\cht{z}{\dual{A}}}}}
\]
\[
\infer[\rlabel{\rname{\ref*{rule:axiom}$_2$}}{rule:cut-backward}]
	{\pttrs
	  {\infer[\ref{rule:cut}] 
     {\judge{\res{w}{x}{P}}{\Gamma \pp \gamma, \cht{y}{A}}}
     {\judge{P}{\Gamma\pp \gamma, \cht{w}{A}\pp \cht{x}{\dual{A}},\cht{y}{A}}}}
	  {\tau}
	  {\judge{Q\{y/w\}}{\Gamma \pp \gamma, \cht{y}{\dual{A}}}}
  }{\pttrs
		{\judge{P}{\Gamma \pp \gamma, \cht{w}{A} \pp \cht{x}{\dual{A}},\cht{y}{A}}}
		{\lforward{x}{y}{A}}
		{\judge{Q}{\Gamma \pp \gamma,\cht{w}{A}}}}
\]
\end{highlight}

\subsection{Exponentials}
In \CT, clients can interact with servers in three ways: they request a service, dispose of a 
server, or request server duplication. Requesting a service is modelled by the following two 
(dual) axioms and rule (for cut elimination).
\begin{highlight}
\[\pttrs
	{\infer[\ref{rule:?}]
		{\judge{\clientuse x{y}P}{\gamma, \cht{x}{\query A}}}
		{\judge{P}{\gamma, \cht{y}{A}}}}
	{\lclientuse x{y}A}
	{\judge{P}{\gamma, \cht{y}{A}}}
\]
\[\pttrs
	{\infer[\ref{rule:!}]
		{\judge{\server x{y}P}{\query\gamma, \cht{x}{\bang A}}}
		{\judge{P}{\query\gamma, \cht{y}{A}}}}
	{\lserveruse x{y}A}
	{\judge{P}{\query\gamma, \cht{y}{A}}}
\]
\[\infer[\rlabel{\ref*{rule:!}\ref*{rule:?}}{rule:cut-?}]
{\vpttrs
{\infer[\ref{rule:cut}]
{\judge
	{\res{x}{y}{P}}
	{\Gamma \pp \gamma, \delta}}
{\judge
	{P}
	{\Gamma \pp \gamma, \cht{x}{\bang A} \pp \delta,\cht{y}{\query \dual A}}}}
{\tau}
{\infer[\ref{rule:cut}]
{\judge
	{\res{x'}{y'}{Q}}
	{\Gamma \pp \gamma, \delta}}
{\judge
	{Q}
	{\Gamma \pp \gamma, \cht{x'}{A} \pp \delta,\cht{y'}{\dual A}}}}}
{\vpttrs
	{\judge
		{P}
		{\Gamma \pp \gamma, \cht{x}{\bang A} \pp \delta,\cht{y}{\query \dual A}}}
	{\lsync{\lserveruse{x}{x'}{A}}{\lclientuse{y}{y'}{\dual{A}}}}
	{\judge{Q}
		{\Gamma \pp \gamma, \cht{x'}{A} \pp \delta', \cht{y'}{\dual A}}}}
\]
\end{highlight}
Server disposal is captured by the following dual axioms and rule.
\begin{highlight}
\[\pttrs
	{\infer[\ref{rule:weaken}]
		{\judge{\clientdisp xP}{\gamma, \cht{x}{\query A}}}
		{\judge{P}{\gamma}}}
	{\lclientdisp xA}
	{\judge{P}{\gamma}}
\]
\[
	{\pttrs
		{\infer[\ref{rule:!}]
			{\judge{\server xyP}{\query \gamma, \cht{y}{\bang A}}}
			{\judge{P}{\query \gamma, \cht{y}{A}}}}
		{\lserverdisp x{A}}
		{\infer[\ref{rule:mix_0}]{\judge{\nil}{\cdot}}{\vphantom{\pp}}}}
\]
\[\infer[\rlabel{\ref*{rule:!}\ref*{rule:weaken}}{rule:cut-weaken}]
{\vpttrs
{\infer[\ref{rule:cut}]
{\judge
	{\res{x}{y}{P}}
	{\Delta \pp \query \gamma, \delta}}
{\judge
	{P}
	{\Delta \pp \query \gamma, \cht{x}{\bang A} \pp \delta,\cht{y}{\query \dual A}}}}
{\tau}
{\infer=[\ref{rule:weaken}]
	{\judge{\clientdisp{z_1}{\dots\clientdisp{z_n}{Q}}}{\Delta \pp\query \gamma,\delta}}
	{\judge
		{Q}
		{\Delta \pp \delta}}}}
{	{\vpttrs[b]
	{\judge
		{P}
		{\Delta \pp \query \gamma, \cht{x}{\bang A} \pp \delta,\cht{y}{\query \dual A}}}
	{\lsync{\lserverdisp{x}{A}}{\lclientdisp{y}{\dual{A}}}}
	{\judge{Q}{\Delta \pp \delta}}}
	&{\phl{\{z_1,\dots,z_n\} = \fn(P)\setminus\fn(Q)}}}
\]
\end{highlight}
Finally, server duplication is defined by the following dual axioms and rule.
We use the convention of adding a prime to indicate a homomorphic copy where every free name $z$ is replaced with $z'$, likewise for (hyper)sequents.
\begin{highlight}
\[\pttrs
	{\infer[\ref{rule:contract}]
		{\judge{\clientspawn x{x'}P}{\gamma, \cht{x}{\query A}}}
		{\judge{P}{\gamma, \cht{x}{\query A},\cht{x'}{\query A}}}}
	{\lclientspawn x{x'}A}
	{\judge{P}{\gamma, \cht{x}{\query A}, \cht{x'}{\query A}}}
\]
\[
{\vpttrs
	{\infer[\ref{rule:!}]
		{\judge{\server x{y}P}{\query\gamma, \cht{x}{\bang A}}}
		{\judge{P}{\query\gamma, \cht{y}{A}}}}
	{\lserverspawn x{x'}A}
	{
		\infer[\ref{rule:mix}]
		{\judge
			{\server{x}{y}{P} \pp \server{x'}{y}{P'}}
			{\query\gamma,\cht{x}{\bang A} \pp
				\query\gamma',\cht{x'}{\bang A}}}
		{
			\infer[\ref{rule:!}]
			{\judge{\server x{y}P}{\query\gamma, \cht{x}{\bang A}}}
			{\judge{P}{\query\gamma, \cht{y}{A}}}
			&
			\infer[\ref{rule:!}]
			{\judge{\server {x'}{y}{P'}}{\query\gamma', \cht{x'}{\bang A}}}
			{\judge{P'}{\query\gamma', \cht{y}{A}}}
		}
	}}
\]

\[\infer[\rlabel{\ref*{rule:!}\ref*{rule:contract}}{rule:cut-contract}]
{\vpttrs
{\infer[\ref{rule:cut}]
{\judge
	{\res{x}{y}{P}}
	{\Delta \pp \query \gamma, \delta,\varepsilon}}
{\judge
	{P}
	{\Delta \pp \query \gamma, \cht{x}{\bang A} \pp \delta,\varepsilon,\cht{y}{\query \dual A}}}}
{\tau}
{\infer=[\ref{rule:contract}]
{\judge
	{\clientspawn{z_1}{z_1'}{\dots\clientspawn{z_n}{z_n'}
		{\res{x}{y}{\res{x'}{y'}{Q}}}}}
	{\Delta \pp \query \gamma, \delta,\varepsilon}}
{\infer[\ref{rule:cut}]
{\judge
	{\res{x}{y}{\res{x'}{y'}{Q}}}
	{\Delta \pp \query\gamma,\query\gamma',\delta,\varepsilon}}
{\infer[\ref{rule:cut}]
{\judge
	{\res{x'}{y'}{Q}}
	{\Delta
	\pp \query\gamma,\cht{x}{\bang A} 
	\pp \query \gamma',\varepsilon,\delta,\cht{y}{\query \dual A}}}
{\judge{Q}
	{\Delta \pp \query\gamma,\cht{x}{\bang A} \pp \query \gamma' \cht{x'}{\bang A} 
	\pp \delta,\cht{y}{\query \dual A} \pp \varepsilon,\cht{y'}{\query \dual A}}}
}}}}
{\deduce
	{\vpttrs
	{\judge
		{P}
		{\Delta \pp \query\gamma, \cht{x}{\bang A} \pp \delta,\varepsilon,\cht{y}{\query \dual A}}}
	{\lsync{\lserverspawn{x}{x'}{A}}{\lclientspawn{y}{y'}{\dual{A}}}}
	{\judge{Q}
		{\Delta
		\pp \query\gamma,\cht{x}{\bang A} \pp \query \gamma', \cht{x'}{\bang A} 
		\pp \delta,\cht{y}{\query \dual A} \pp \varepsilon,\cht{y'}{\query \dual A}}}
		}
	{\phl{\{z_1,\dots,z_n\} = \fn(Q)\setminus\fn(P)}}}
\]
\end{highlight}
\looseness=-1
One may wonder why server dependencies are handled by \cref{rule:cut-weaken,rule:cut-contract} and not by axioms for \ref{rule:!}, there are two main reasons:
The first is that in this way resource use is observable from labels as well as the type in the underlying proof-transformation.
The second is adherence with \CP where the contraction rule is applied outside cuts (\cf \cite[Fig.~3]{W14}) meaning that dependencies are (silently) duplicated outside the restriction where they are needed---as in \cref{rule:cut-contract}.
We observe however, that it is possible to move resource management to axioms for server disposal and spawning once rule for external (\ie at the hypersequent level) contraction is added to the type system (external weakening is derivable in this case)

\subsection{Polymorphism}
Polymorphism is achieved by communicating types, according to the following transition axioms and 
rule.
\begin{highlight}
\[
\pttrs
	{\infer[\ref{rule:exists}]
		{\judge{\sendtype xAP}{\gamma, \cht{x}{\exists X.B}}}
		{\judge{P}{\gamma, \cht{x}{B\{A/X\}}}}}
	{\lsendtype xA}
	{\judge{P}{\gamma, \cht{x}{B\{A/X\}}}}
\]
\[
\pttrs
	{\infer[\ref{rule:forall}]
		{\judge{\recvtype yYP}{\gamma, \cht{y}{\forall Y.B}}}
		{\judge{P}{\gamma, \cht{y}{B}}}} 
	{\lrecvtype yA}
	{\judge{P}{\gamma, \cht{x}{B\{A/Y\}}}}
\]
\[
\infer[\rlabel{\ref*{rule:exists}\ref*{rule:forall}}{rule:cut-exists}]
{\vpttrs
	{\infer[\ref{rule:cut}] 
		{\judge{\res{x}{y}{P}}{\Gamma \pp \gamma,\delta}}
		{\judge{P}{\Gamma\pp \gamma,\cht{x}{\exists X.B}  \pp \delta, \cht{y}{\forall Y.C}}}}
	  {\tau}
	{\infer[\ref{rule:cut}]
		{\judge{\res{x}{y}{Q}}{\Gamma \pp \gamma,\delta}}
		{\judge{Q}{\Gamma\pp \gamma,\cht{x}{B\{A/X\}} \pp \delta, \cht{y}{C\{\dual{A}/Y\}}}}}}
{\vpttrs
	{\judge{P}{\Gamma\pp\gamma,\cht{x}{\exists X.B}  \pp \delta, \cht{y}{\forall Y.C}}}
	{\lsync{\lsendtype xA}{\lrecvtype y{\dual A}}}
	{\judge{Q}{\Gamma \pp \gamma,\cht{x}{B\{A/X\}} \pp \delta,\cht{y}{C\{\dual{A}/Y\}}}}}
\]
\end{highlight}

\section{Metatheory: Subject Reduction, Progress, Termination}

The transition rules of \CT are derived from proof transformations that preserve provability.
Observe that all rules for $\tau$-transitions enjoy type preservation: preserve the types in the 
judgement in the conclusion remain unchanged. All the other transitions, for observable actions, 
preserve provability: types change depending on the action(s), but the concluding judgement remains 
valid.
Thus we immediately obtain the following theorem.
\begin{theorem}[Subject Reduction]
\label{thm:subjred}
Let $\judge{\nohl P}{\nohl\Gamma}$ and $P \lto\alpha Q$. Then:
\begin{itemize} 
\item $\alpha = \tau$ if and only if $\judge{\nohl Q}{\nohl \Gamma}$;
\item $\alpha \neq \tau$ if and only if $\judge{\nohl Q}{\nohl \Delta}$ for some $\Delta \neq 
\Gamma$.
\end{itemize}
\end{theorem}

\Cref{thm:subjred} formalises that $\tau$-transitions of a process $P$ have no dependencies on the 
context, since they do not influence the type of $P$. This matches the intuition of LTS semantics 
for the $\pi$-calculus, where $\tau$-transitions capture internal ``unobservable'' moves. In \CT, 
performing unobservable moves coincides with type-preserving proof transformations.

\CT also enjoys progress, in the sense that well-typed programs are either terminated or admit a 
transition.
\begin{theorem}[Progress]
If $\judge{\nohl P}{\nohl\Gamma}$, then either $P$ is terminated or there exist $\alpha$ and $Q$
such that $P\lto\alpha Q$.
\end{theorem}

\looseness=-1
\CT enjoys weak termination. The only source of divergence in \CT is the term for 
replicated processes, more precisely when a server is not paired (\rname{Cut}) with its 
client. Then, we can observe an infinite number of replications.
However, we can also decide at any moment to choose the transition that visibly disposes of the 
server. In the following, $\lto{\til\alpha}^{\ast}$ is the transitive closure of $\lto\alpha$, \ie, 
$P \lto{\til\alpha}^{\ast} Q$ means $P_1 \lto{\alpha_1} \cdots \lto{\alpha_n} Q$ for some $n$ 
(possibly zero, in which case $P=Q$ and $\til\alpha$ is empty).
\begin{theorem}[Weak Termination]
If $\judge{\nohl P}{\nohl\Gamma}$, then there exist $\til\alpha$ and $Q$ such that 
$P\lto{\til\alpha} Q$ and $Q$ is terminated.
\end{theorem}

\section{Related Work}

Since its inception, linear logic was described as the logic of concurrency \citep{G87}.
Correspondences between the proof theory of linear logic and variants of the $\pi$-calculus emerged 
soon afterwards \citep{A94,BS94}, by interpreting linear logic propositions as types for channels. 
Later, linearity inspired also the seminal theories of linear types for the $\pi$-calculus 
\cite{KPT99} and session types \cite{HVK98}. Even though the two theories do not use exactly linear 
logic, the work by \citet{DGS17} shows that the link is still strong enough that session types can 
be encoded into linear types.

It took more than ten years for a formal correspondence between linear logic and (a variant of) 
session types to emerge, with the seminal paper by \citet{CP10}. This then inspired the development 
of Classical Processes (\CP) by \citet{W14}, which we have already discussed in the Introduction. 
We have extended this line of work to labelled transition systems.

The idea of extending linear logic to hypersequents for typing processes is not new.
Specifically, in \cite{CMS17}, the multiplicative-additive fragment of intuitionistic linear logic 
is extended with hypersequents to type choreographies (global descriptions of process 
communications). Differently, \CT is based on \CLL, uses hypersequents to type processes, and deals 
also with exponentials and polymorphism. The major difference is that the rules for 
manipulating hypersequents are different \wrt \cite{CMS17}. In particular, hypersequents can be 
formed in \cite{CMS17} only when sequents share resources (compare to our \rname{Mix}, which 
requires the opposite), and these resource sharings are then explicitly tracked using an additional 
connection modality (which is not present in \CT).
As a consequence, the work in \cite{CMS17} does not support an LTS for the same reasons that we 
discussed for \CP in the Introduction. Adding a connection modality to \CT might be interesting for 
providing an LTS semantics to the choreographies studied in \cite{CMS17}.

\section{Conclusions}
\label{sec:conclusion}

\looseness=-1
We believe that \CT might spark a new line of research on the observable behaviour of 
processes typed with the theory of linear logic.
An immediate direction would be studying behavioural theories for \CT.
For example, standard bisimulation techniques have not been applied to these calculi yet, and 
instead new definitions of behavioural equivalences based on logical relations or denotational 
semantics have been devised for linear logic proof terms \cite{PCPT14,A17}.
The main reason is that linear logic exhibits more concurrent behaviour than usual, which makes contextual equivalence more coarse than in the $\pi$-calculus. For example, the processes 
$\nohl{\wait{x}{\wait{y}{P}}}$ and $\nohl{\wait{y}{\wait{x}{P}}}$ would be distinguished by standard (strong) bisimulation, but they are 
contextually equivalent because typing enforces $x$ and $y$ to be connected to separate parallel processes, as observed in \cite{A17}. We believe that adding delayed actions (or variations thereof) bridge this gap, \eg, these two processes in particular would now be equated by bisimulation. A thorough investigation of the relation between bisimulation and contextual equivalence in \CT is left to future work.

\end{document}